# A Decoupled Scheme to Solve the Mass and Momentum Conservation Equations of the Improved Darcy-Brinkman-Forchheimer Framework in Matrix Acidization*

Yuanqing Wu[1], Jisheng Kou[2#], Yu-Shu Wu[3], Shuyu Sun[1], Yilin Xia[1]

[1]*College of Mathematics and Statistics, Shenzhen University, Shenzhen, Guangdong 518048, China*

[2]*School of Civil Engineering, Shaoxing University, Shaoxing, Zhejiang 312000, China*

[3]*Department of Petroleum Engineering, Colorado School of Mines, 1600 Arapahoe Street, Goden, CO 80401, USA*

## Abstract

Matrix acidization simulation is a challenging task in the study of flows in porous media, due to the changing porosity in the procedure. The improved DBF framework is one model to do this simulation, and its numerical scheme discretises the mass and momentum conservation equations together to form a pressure-velocity linear system. However, this linear system can only be solved by direct solvers to solve for pressure and velocity simultaneously, since zeros appear in the diagonal of the coefficient matrix. Considering the large-scale attribute of matrix acidization simulation, the solving time of direct solvers is not intolerant. Thus, a decoupled scheme is proposed in this work to decouple the coupled pressure-velocity linear system into two independent linear systems: one is to solve for pressure, and the other one is to solve for velocity. Both of the new linear systems can be solved by parallel and iterative solvers, which guarantees the large-scale simulation can be finished in a reasonable time period. A numerical experiment is carried out to demonstrate the correctness of the decoupled scheme and its higher computing efficiency.

## 1. Introduction

In petroleum engineering, matrix acidization is a kind of operation, in which the oil production channel can be opened up again by the injection of acid flows, after the acid flows dissolve the depositions such as mud and rocks. However, the numerical simulation of matrix acidization is not an easy task in the flow study of porous media. Its key characteristic is the change of porosity with the progress of acidization procedure, which brings a challenge to the simulation task. In order to reflect the change of porosity accurately, the Darcy-Brinkman-Forchheimer (DBF) framework and its derivations are provided by Wu et al ([1]-[5]). In this framework, the behaviour of acid flows in the matrix is decided not only by Darcy's law, but also by the viscous effect and inertial force, which is represented by the adding of Brinkman term and Forchheimer term to the momentum conservation equation of the two-scale model [6], respectively. After this, the DBF framework is named. The meaning of all kinds of notations in the following discussions is given in Table 1.

Table 1 Nomenclature

| Notation | Meaning | Notation | Meaning |
|---|---|---|---|
| $\boldsymbol{p}$ | pressure | $t$ | time |
| μ | fluid viscosity | $\boldsymbol{g}$ | gravity vector |
| $\boldsymbol{K}$ | permeability value | $C_f$ | cup-mixing concentration of the acid |

* This work is supported by Peacock Plan Foundation of Shenzhen (No. 000255), National Natural Science Foundation of China (No. 11601345) and Natural Science Foundation of SZU (No. 2017059).

# Corresponding Author. E-mail address: koujisheng@163.com.

| | | | |
|---|---|---|---|
| $\boldsymbol{u}$ | velocity vector | $\boldsymbol{D_e}$ | effective dispersion tensor |
| $\boldsymbol{\phi}$ | porosity | $k_c$ | local mass-transfer coefficient |
| $\rho_f$ | mass density of the fluid | $a_v$ | interfacial surface area per unit volume |
| $\boldsymbol{F}$ | Forchheimer coefficient | $C_s$ | concentration of the acid at the fluid-solid interface |
| | | $\varepsilon$ | pseudo-compressibility factor |

In the former literature, the mass conservation equation and momentum conservation equation of the improved DBF framework [5] (one of the derivations of the DBF framework) are combined together to form an equation system to solve for the pressure field and velocity field simultaneously. With the finite difference method in space and semi-implicit method in time, this equation system can be discretized to a linear system. Furthermore, such kind of linear system can be solved appropriately with the serial and direct solvers such as LAPACK [7] and UMFPACK [8], if the linear system is not too large. However, as mentioned above, the key characteristic of matrix acidization is the change of porosity, which requires a fine enough simulation grid to capture the structure of porosity well, i.e., the linear system will become very large. Unfortunately, the computation of the serial and direct solvers is very time wasting in such condition, which leads to the intolerable running time of matrix acidization simulation. One plausible solution is appealing to the parallel and iterative solvers such as HYPRE [9]. However, such kind of solvers meet one main trouble when solving the linear system, which can be seen as below. The momentum conservation equation of the improved DBF framework is

$$\rho_f \frac{\partial}{\partial t}\left(\frac{\boldsymbol{u}}{\phi}\right) + \rho_f \frac{\boldsymbol{u}}{\phi} \cdot \nabla \frac{\boldsymbol{u}}{\phi} = -\nabla p - \frac{\mu}{K}\boldsymbol{u} + \nabla \cdot \mu \nabla \frac{\boldsymbol{u}}{\phi} - \frac{\rho_f F}{\sqrt{K}}|\boldsymbol{u}|\boldsymbol{u} + \rho_f \boldsymbol{g}. \tag{1}$$

The mass conservation equation of the improved DBF framework is

$$\frac{\partial \phi}{\partial t} + \nabla \cdot \boldsymbol{u} = 0. \tag{2}$$

After discretization, the linear system $\boldsymbol{Ax} = \boldsymbol{b}$ is shown in Figure 1. If the 2D domain is considered, the discretized $x$-momentum conservation equations, the discretized $y$-momentum conservation equations and the discretized mass conservation equations are ranked one by one from the top to the bottom of the linear system. Accordingly, the $x$-velocity field $\boldsymbol{U}$, the $y$-velocity field $\boldsymbol{V}$ and the pressure field $\boldsymbol{P}$ are put into the unknown vector $\boldsymbol{x}$ one by one. It is easy to see that the entries of the diagonal at the right-bottom corner of the coefficient matrix $\boldsymbol{A}$ are zeros, which is the reason that the iterative solvers cannot solve the linear system. Some works [2] suggest adding a compressibility term to the mass conservation equation, and then Equation (2) is changed to

$$\varepsilon \frac{\partial p}{\partial t} + \frac{\partial \phi}{\partial t} + \nabla \cdot \boldsymbol{u} = 0. \tag{3}$$

After such processing, the entries of the diagonal of the coefficient matrix $\boldsymbol{A}$ are non-zeros, and therefore the linear system can be solved by iterative solvers. However, the introduction of the compressibility term changes the attribute of the flows, which should be incompressibility in nature. A trade-off is to set the pseudo-compressibility factor $\varepsilon$ to be a very small positive number.

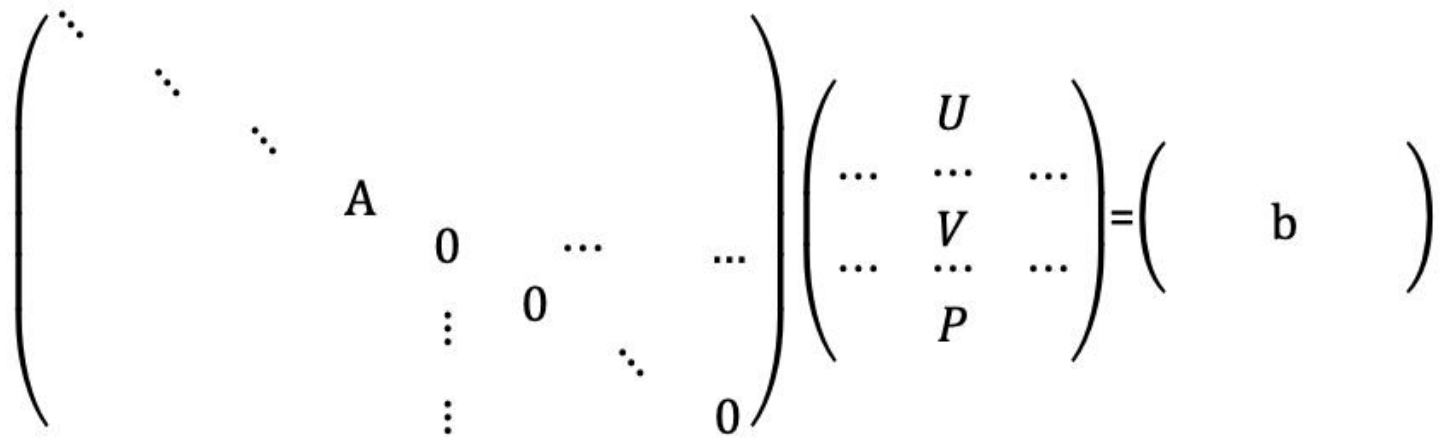

Figure 1 The pressure-velocity linear system after discretization.

## 2. Decoupled Scheme

Although the trade-off strategy fixes the solvability issue of the linear system with iterative solvers, it cannot manifest the incompressibility attribute of the flows in matrix acidization. Thus, a decoupled scheme is suggested to solve the problem. In this scheme, the Darcy-Forchheimer equation is solved firstly, which can be shown as below

$$\frac{\mu}{K}\boldsymbol{u}_D^{n+1} + \frac{\rho_f F}{\sqrt{K}}|\boldsymbol{u}^n|\boldsymbol{u}_D^{n+1} = -\nabla p^{n+1} + \rho_f \boldsymbol{g}, \quad (4)$$

in which $\boldsymbol{u}_D$ is an intermediate velocity, and the superscripts represent the time step. By Equation (4), $\boldsymbol{u}_D$ can be solved as

$$\boldsymbol{u}_D^{n+1} = \left(\frac{\mu}{K} + \frac{\rho_f F}{\sqrt{K}}|\boldsymbol{u}^n|\right)^{-1}\left(-\nabla p^{n+1} + \rho_f \boldsymbol{g}\right). \quad (5)$$

In this condition, the mass conservation equation can be written as

$$\frac{\partial \phi}{\partial t} + \nabla\cdot \boldsymbol{u}_D^{n+1} = 0. \quad (6)$$

Then, by substituting Equation (5) to (6), the pressure equation can be obtained as

$$\frac{\partial \phi}{\partial t} + \nabla\cdot \left(\frac{\mu}{K} + \frac{\rho_f F}{\sqrt{K}}|\boldsymbol{u}^n|\right)^{-1}\left(-\nabla p^{n+1} + \rho_f \boldsymbol{g}\right) = 0. \quad (7)$$

This equation is an elliptic equation with the unknown $p^{n+1}$, and therefore all the entries of the diagonal of the coefficient matrix are non-zeros, which guarantees its solvability by iterative solvers. Not only that, the incompressibility attribute of the flows is also guaranteed. However, the intermediate velocity $\boldsymbol{u}_D$ is computed only from the Darcy term and Forchheimer term, and therefore the Brinkman correction should be included as

$$\frac{\rho_f}{\Delta t}\left(\frac{\boldsymbol{u}^{n+1}}{\phi^{n+1}} - \frac{\boldsymbol{u}^n}{\phi^n}\right) + \left(\frac{\mu}{K} + \frac{\rho_f F}{\sqrt{K}}|\boldsymbol{u}^n|\right)\left(\boldsymbol{u}^{n+1} - \boldsymbol{u}_D^{n+1}\right) + \rho_f \nabla\cdot\left(\frac{\boldsymbol{u}^n}{\phi^n}\otimes\frac{\boldsymbol{u}^{n+1}}{\phi^{n+1}}\right) - \mu\nabla^2\frac{\boldsymbol{u}^{n+1}}{\phi^{n+1}} = 0. \quad (8)$$

Equation (8) is another elliptic equation with the unknown $\boldsymbol{u}^{n+1}$. With the same reason as above, it can be solved by iterative solvers. Moreover, it is easy to see that if Equation (4) is substituted into (8), the discretized Equation (1) can be achieved, which means that $\boldsymbol{u}^{n+1}$ in Equation (8) also satisfy (1). This is a demonstration that the decoupled scheme has the same velocity solution as the improved DBF framework. In the decoupled scheme, the pressure computed from Equation (7) is an intermediate pressure instead of the true pressure, since it is derived from the Darcy-Forchheimer equation (4). The intermediate pressure is enough to compute the velocity, and the true pressure will not influence the computing of the other variables. Thus, the true pressure will not be computed. In sum, the decoupled scheme has the same solution as the improved DBF framework, except for the pressure.

Thus, by the mathematical derivation above, the coupled pressure-velocity linear system of the improved DBF framework is decoupled into two linear systems: one is to solve for the pressure field, and the other one is to solve for the velocity field. Since both of the two linear systems are derived from the elliptic equations, they can be solved by iterative solvers. Furthermore, if parallel and iterative solvers are used, the large-scale simulation of matrix acidization can be finished in a reasonable time period by a parallel simulator. Meanwhile, since the decoupled scheme has the same solution (except pressure) as the improved DBF framework, the incompressibility attribute of flows in matrix acidization is maintained.

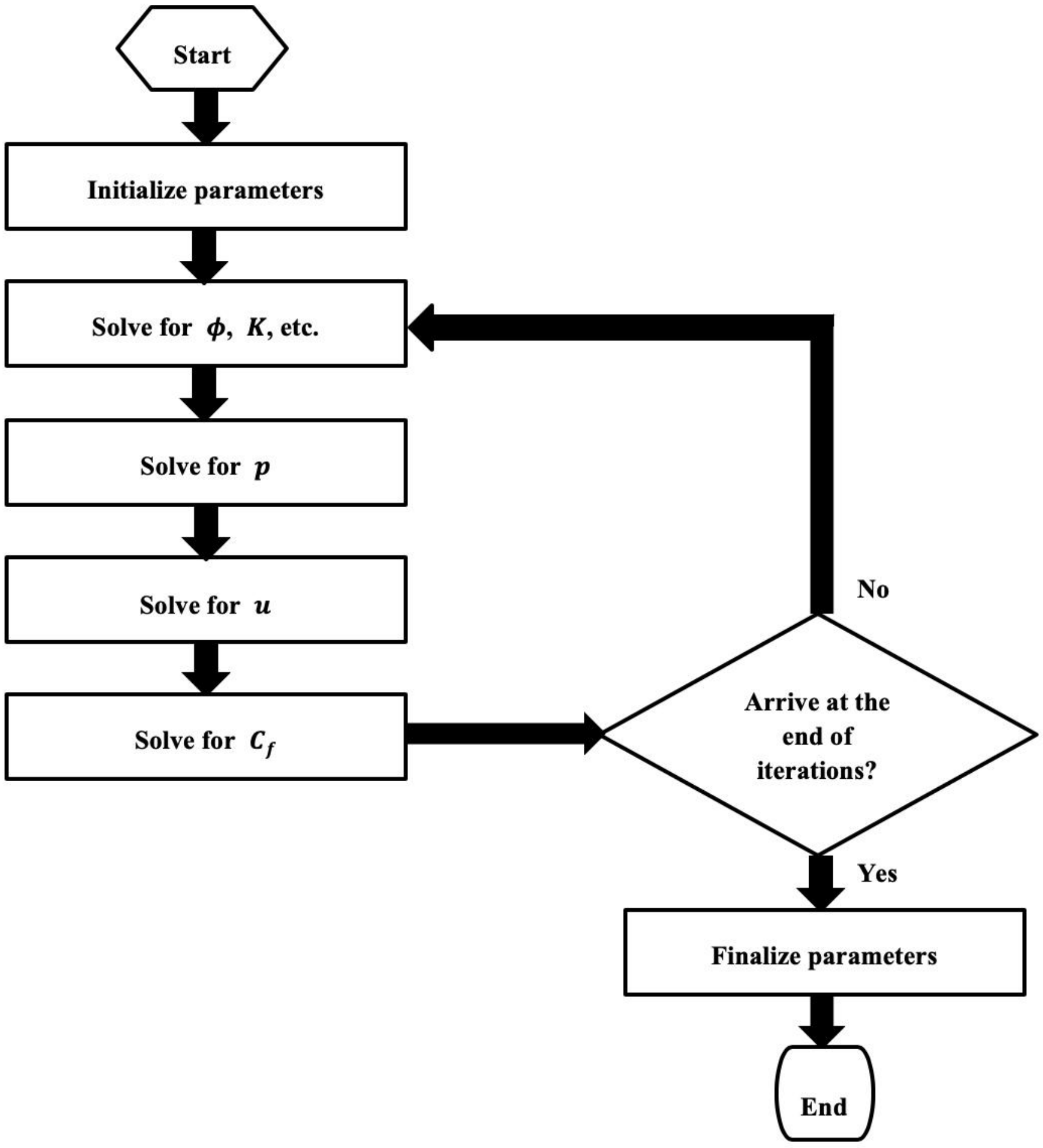

Figure 2 The flowchart of the decoupled scheme.

## 3. Experiment

A numerical experiment is carried out in this section to demonstrate the reliability and applicability of the decoupled scheme. For the purpose of comparison with the improved DBF framework, the concentration equation (9) is introduced to finish the description of the improved DBF framework.

$$\frac{\partial(\phi C_f)}{\partial t} + \nabla\cdot(\boldsymbol{u}C_f) = \nabla\cdot(\phi \boldsymbol{D}_e\cdot\nabla C_f) - k_c a_v(C_f - C_s). \quad (9)$$

The other equations of the improved DBF framework will not be given here again, and they can be referenced in ([1][5]). With the same philosophy as before, one linear system with the unknown $C_f^{n+1}$ is formed, after Equation (9) is discretized. Since Equation (9) is an elliptic equation, the linear system can be solved by iterative solvers. Until now, there are three linear systems in the decoupled scheme, and all of them can be solved by iterative solvers. The flowchart of the decoupled scheme can be shown in Figure 2.

Although the parallel simulator of matrix acidization was realized in the existing works, and the parallel solver such as MUMPS [10] was applied in the simulator, the merits of parallelism are not fully exploited considering MUMPS is a direct solver. With the decoupled scheme, parallel and iterative solvers such as HYPRE can be applied in the simulator, and much more acceleration can be gained. In the following experiment, three tasks will be done:

1. Two single processors with the solver UMFPACK run on the improved DBF framework and the decoupled scheme, respectively, to investigate the correctness of the decoupled scheme.
2. The solver time of both the improved DBF framework and decoupled scheme is given to compare the computing efficiency of them.
3. The parallel and iterative solver HYPRE is applied to the decoupled scheme to evaluate the parallel performance of the simulator.

In this experiment, matrix acidization is done in a rectangular domain with the size of 0.1-m by 0.04-m. Acid flow is injected into the domain from the left side with the velocity of $4.17\times 10^{-6}$ m/s. A pressure of $1.52\times 10^{7}$ Pa is imposed on the right side of the domain. The up and down boundaries of the domain are closed. The concentration of the injected acid flow is $5.0\times 10^{2}$ mol/m$^2$. The average initial porosity of the matrix is 0.18, and the average initial permeability of the matrix is $9.869\times 10^{-16}$ m$^2$. The simulator is written in FORTRAN 90 and MPI, and runs on macOS Catalina with four cores. If a grid of 180×72 cells is imposed on the domain, and the time step is 10 s, the simulation results are shown as below.

Table 2 Results comparison

| | **Improved DBF framework** | **Decoupled scheme** |
|---|---|---|
| **Pore volume to breakthrough** | 4.866 | 4.848 |
| **Average porosity** | 0.593 | 0.592 |
| **Average concentration** | 268.325 mol/m$^2$ | 268.550 mol/m$^2$ |
| **Solver time** | 40177.5 s | 23523.5 s |

For Task 1, the key parameters at breakthrough are compared in Table 2. It can be seen that the pore volume to breakthrough, average porosity and average concentration in the domain are more or less the same for the improved DBF framework and decoupled scheme. Moreover, the porosity profile, concentration profile and the streamlines are also compared in Figure 3. From the table and the figure, it can be said that both the improved DBF framework and the decoupled scheme simulate very similar results. In such condition, the correctness of the decoupled scheme is demonstrated by the experiment.

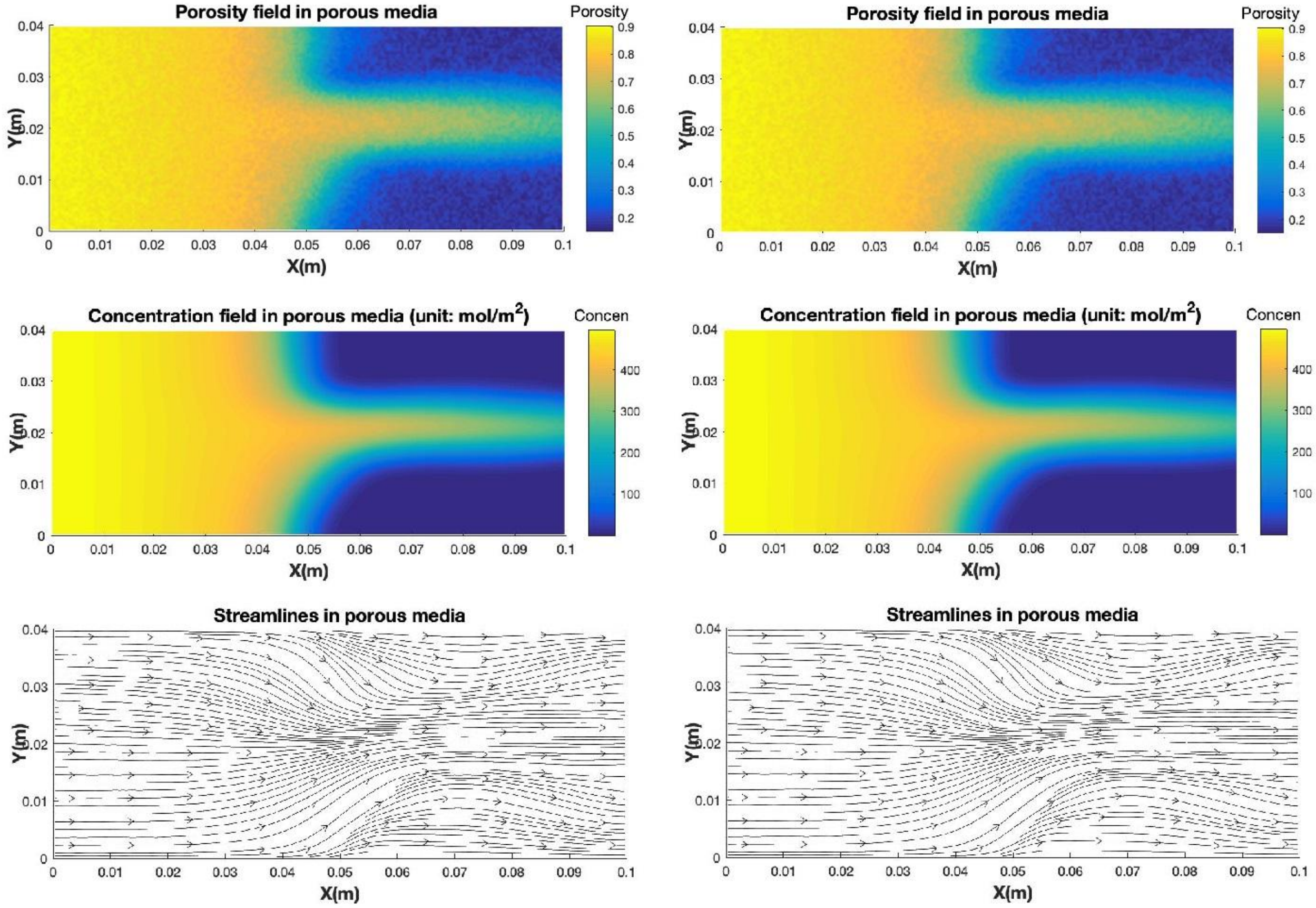

Improved DBF framework Decoupled scheme

Figure 3 The porosity profile, concentration profile and streamlines in the domain for the improved DBF framework (left column) and decoupled scheme (right column).

For Task 2, it can be known from the last row of Table 2 that the solver time of the decoupled scheme is much less than the counterpart of the improved DBF framework, which demonstrates that the decoupled scheme has much higher computing efficiency than the improved DBF framework.

For Task 3, since there are only four cores in our Mac, the simulator will use one, two, and four processors to run, respectively. GMRES [11] in HYPRE is used to solve all the three linear systems. The solver time is 2267.0 s, 1141.4 s and 600.7 s, respectively, for the processors above. From the result, it can be known that the parallel and iterative solver HYPRE has a good scalability on the decoupled scheme.

## 4. Conclusion

Accurate simulation of matrix acidization requires the change of porosity be well captured by the simulator, which can be guaranteed by the use of fine enough grids. However, such kind of grids makes this simulation fall in the field of large-scale simulation, and the application of parallel and iterative solvers becomes mandatory. The former numerical scheme of the improved DBF framework generated a coupled pressure-velocity linear system, and both pressure and velocity were solved together.

Considering that zeros appear in the diagonal of the coefficient matrix, the coupled linear system can only be solved by direct solvers, which leads to huge running time of the large-scale simulation as mentioned above. Thus, in this work, a decoupled scheme is provided to decouple the coupled pressure-velocity linear system into two linear systems: one is to solve for pressure, and the other one is to solve for velocity. By this way, both of the new linear systems can be solved by parallel and iterative solvers such as HYPRE. As a result of that, the large-scale simulation can be finished in a reasonable time period. We also demonstrate mathematically that the decoupled scheme has the same simulation results as the coupled scheme of the improved DBF framework, except for the pressure, which is also proved by the numerical experiment. Besides that, the numerical experiment also shows the higher computing efficiency of the decoupled scheme, and the good scalability of HYPRE on the decoupled scheme.

The next step of this work is to derive the decoupled scheme for the thermal DBF framework [12], which is an expansion of the improved DBF framework by adding the energy conservation equation. However, this kind of derivation is not straightforward. As mentioned above, the pressure $p$ derived from the decoupled scheme is not the true pressure, which leads to that the pressure $p$ cannot be used directly in the energy conservation equation of the thermal DBF framework. Some kind of correction form of the pressure $p$ should be considered.